\documentclass[twocolumn,showpacs,preprintnumbers,amsmath,amssymb,superscriptaddress,prl,aps]{revtex4}

\usepackage{graphicx}
\usepackage{dcolumn}
\usepackage{bm}

\begin{document}


\title{Limits on the WIMP-nucleon interactions with CsI(Tl) crystal detectors}

\affiliation{DMRC and Department of Physics and Astronomy, Seoul National University, Seoul, Korea}
\affiliation{Department of Physics, Kyungpook National University, Daegu, Korea}
\affiliation{Department of Physics, Sejong University, Seoul, Korea}
\affiliation{Department of Science Education, Ewha Womans University, Seoul, Korea}
\affiliation{Department of Physics, Younsei University, Seoul, Korea}
\affiliation{Department of Engineering Physics, Tsinghua Universuty, Beijing, China}

\author{H.S. \surname{Lee}}
\affiliation{DMRC and Department of Physics and Astronomy, Seoul National University, Seoul, Korea}
\author{H.C. Bhang}
\affiliation{DMRC and Department of Physics and Astronomy, Seoul National University, Seoul, Korea}
\author{J.H. Choi}
\affiliation{DMRC and Department of Physics and Astronomy, Seoul National University, Seoul, Korea}
\author{H. Dao}
\affiliation{Department of Engineering Physics, Tsinghua University, Beijing, China}
\author{I.S. Hahn}
\affiliation{Department of Science Education, Ewha Womans University, Seoul, Korea}
\author{M.J. Hwang}
\affiliation{Department of Physics, Younsei University, Seoul, Korea}
\author{S.W. Jung}
\affiliation{Department of Physics, Kyungpook National University, Daegu, Korea}
\author{W.G. Kang}
\affiliation{Department of Physics, Sejong University, Seoul, Korea}
\author{D.W. Kim}
\affiliation{DMRC and Department of Physics and Astronomy, Seoul National University, Seoul, Korea}
\author{H.J. Kim}
\affiliation{Department of Physics, Kyungpook National University, Daegu, Korea}
\author{S.C. Kim}
\affiliation{DMRC and Department of Physics and Astronomy, Seoul National University, Seoul, Korea}
\author{S.K. Kim}
\email[]{skkim@hep1.snu.ac.kr}
\affiliation{DMRC and Department of Physics and Astronomy, Seoul National University, Seoul, Korea}
\author{Y.D. Kim}
\affiliation{Department of Physics, Sejong University, Seoul, Korea}
\author{J.W. Kwak}
\altaffiliation[Current address: ]{National Cancer Center, Ilsan, Korea}
\affiliation{DMRC and Department of Physics and Astronomy, Seoul National University, Seoul, Korea}
\author{Y.J. Kwon}
\affiliation{Department of Physics, Younsei University, Seoul, Korea}
\author{J. Lee}
\altaffiliation[Current address: ]{Department of Physics, Ewha Womans University, Seoul, Korea}
\affiliation{DMRC and Department of Physics and Astronomy, Seoul National University, Seoul, Korea}
\author{J.H. Lee}
\affiliation{DMRC and Department of Physics and Astronomy, Seoul National University, Seoul, Korea}
\author{J.I. Lee}
\affiliation{Department of Physics, Sejong University, Seoul, Korea}
\author{M.J. Lee}
\affiliation{DMRC and Department of Physics and Astronomy, Seoul National University, Seoul, Korea}
\author{S.J. Lee}
\affiliation{DMRC and Department of Physics and Astronomy, Seoul National University, Seoul, Korea}
\author{J. Li}
\affiliation{Department of Engineering Physics, Tsinghua University, Beijing, China}
\author{X. Li}
\affiliation{Department of Engineering Physics, Tsinghua University, Beijing, China}
\author{Y.J. Li}
\affiliation{Department of Engineering Physics, Tsinghua University, Beijing, China}
\author{S.S. Myung}
\affiliation{DMRC and Department of Physics and Astronomy, Seoul National University, Seoul, Korea}
\author{S. Ryu}
\affiliation{DMRC and Department of Physics and Astronomy, Seoul National University, Seoul, Korea}
\author{J.H. So}
\affiliation{Department of Physics, Kyungpook National University, Daegu, Korea}
\author{Q. Yue}
\affiliation{Department of Engineering Physics, Tsinghua University, Beijing, China}
\author{J.J. Zhu}
\affiliation{Department of Engineering Physics, Tsinghua University, Beijing, China}

\collaboration{KIMS Collaboration}
\noaffiliation

\date{\today}

\begin{abstract}
The Korea Invisible Mass Search~(KIMS) experiment presents new 
limits on the WIMP-nucleon cross section using data from an exposure of 3409~kg$\cdot$d taken with low-background CsI(Tl) crystals at Yangyang  
Underground Laboratory. The most stringent limit on the spin-dependent 
interaction for a pure proton case is obtained. The 
DAMA signal region for both spin-independent and spin-dependent 
interactions for the WIMP masses greater than 20 GeV/c$^2$ is excluded 
by the single experiment with crystal scintillators.

\end{abstract}

\maketitle


The existence of dark matter 
has been widely supported by many astronomical observations 
on various scales~\cite{Rcuv}\cite{WMAP}\cite{SDSS}. 
Weakly interacting massive particles~(WIMPs) are 
a good candidate for dark matter well motivated by cosmology 
and supersymmetric models~\cite{jungman}.   
The Korea Invisible Mass Search~(KIMS) experiment has developed low-background CsI(Tl) crystals to detect the signals from the elastic 
scattering of WIMP off the nucleus~\cite{kims07a}\cite{kims02b}\cite{kims05a}. 
Both $^{133}$Cs and $^{127}$I are sensitive to the spin-independent~(SI) 
and spin-dependent~(SD) interactions of WIMPs.
Recently, the role of CsI in the direct search for SD WIMP for pure proton coupling has been pointed out~\cite{girard07}.  
It is worth noting that $^{127}$I is the dominant target for the SI interactions in the DAMA experiment.  
The pulse 
shape discrimination~(PSD) technique allows us to statistically separate 
nuclear recoil~(NR) signals of WIMP interactions from the electron recoil~(ER) 
signals due to the gamma ray background~\cite{kims01}\cite{kims02a}. 

The KIMS experiment is located at the Yangyang Undeground 
Laboratory~(Y2L) at a depth of 700~m under an earth overburden. 
Details of the 
KIMS experiment and the first limit with 237~kg$\cdot$d exposure data can 
be found in the previous publication~\cite{kims06a}. 
Four low-background CsI(Tl) crystals are installed
in the Y2L and operated at a temperature of T~=~$0^\circ$C. 
Throughout the exposure period, the temperature of the detector was kept stable to within $\pm0.1^{\circ}$C.
Green-enhanced photomultiplier tubes~(PMTs) are mounted at both ends of each crystal.
The signals from the PMTs are amplified and recorded by a 
500~MHz FADC. Each event is recorded for a period of 32~$\mu$s. 
Both PMTs on each crystal must have at least two photoelectrons within a 2~$\mu$s window to form an event trigger. 
We obtained 3409~kg$\cdot$d WIMP search data with four crystals, as shown 
in Table~\ref{tab:crystal}.
The energy is calibrated using 59.5~keV gamma 
rays from an $^{241}$Am source. For calibration of the mean 
time, a variable used for the PSD, NR events are 
obtained with small crystals~( 3 cm $\times$ 3 cm $\times$ 3 cm ) using an Am-Be neutron source. Compton scattering events 
taken with the WIMP search crystals using the $^{137}$Cs source are used to determine the mean time 
distribution of the gamma background. Compton scattering events are also taken with the small crystals to verify that the mean time ditributions for both the test crystals and the WIMP search crystals are the same. 
In order to understand the 
nature of the PMT background, 
a dominant background at low energies, acrylic boxes are mounted on the same 
PMTs used for the crystals. 
The data obtained using this setup is used to 
develop the cuts
for the rejection of PMT background.  
\begin{table}
\caption{\label{tab:crystal}Crystals used in this analysis and amount of data for each crystal}
\begin{ruledtabular}
\begin{tabular}{ccc}
Crystal &mass (kg)&data~(kg$\cdot$days)\\
\hline
S0501A& 8.7 & 1147 \\
S0501B& 8.7 & 1030 \\
B0510A& 8.7 & 616 \\
B0510B& 8.7 & 616 \\
\hline
Total& 34.8 & 3409 \\
\end{tabular}
\end{ruledtabular}
\vspace{-10pt}
\end{table}

Since the decay time of the scintillation light in the CsI(Tl) crystal 
is rather long, photoelectrons are well separated at low energies and 
thereby enabling reconstruction of each photoelectron. The time 
distribution of photoelectrons in an event is fitted to a double 
exponential function given by  
$$
f(t) =  {1\over{\tau_f}} \exp \left\{-(t-t_0)\over{\tau_f}\right\} + 
{R\over{\tau_s}}\exp\left\{-(t-t_0)\over{\tau_s}\right\},$$ 
where $\tau_f$ and $\tau_s$ are decay time constants of fast and slow components, respectively, $R$ is ratio between two components, 
and $t_0$ is the time of the first 
photoelectron in the event. 
The mean time~($MT$) of each event is then calculated using these quantities as
$$MT = \int{t\cdot f(t)dt}/\int{f(t)dt}.$$
With this method, an improvement in PSD is achieved over the previous 
analysis where we used a simple mathematical mean~\cite{kims06a}. In order to reject the PMT background, we applied cuts to the fit 
variable, $\tau_f$. The ratio between the maximum log likelihood value of the double exponential fit and that of the single exponential fit is also used to reject the PMT 
background, since PMT background events tend to be shaped as single 
exponential decay. To reject the background that originates from 
the radioactivity of the PMT, the asymmetry between the signals from 
two PMTs is 
applied. Finally events in which signals are recorded in more than one crystal are rejected.
The event selection efficiency was estimated by 
applying the same analysis cuts to the neutron and gamma calibration 
samples. The efficiency depends on the measured energy and ranges from 
30\% at 3 keV to 60\% above 5 keV. 

\begin{figure}
\includegraphics[width=0.44\textwidth]{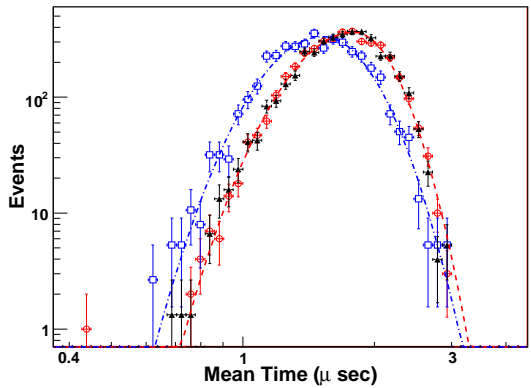}
\caption{\label{MT_comp}(color online). $MT$ distribution of 
NR events~(open squares), ER 
events~(open circles) and WIMP search data~(filled triangles) of S0501A crystal in the 5-6 keV range.
Fitted PDF functions are overlayed. $\chi^2/DOF=$0.8 and 1.3 with $DOF$=38 and 35 for NR and ER events respectively. } 
\vspace{-10pt}
\end{figure}

\begin{figure}
\includegraphics[width=0.44\textwidth]{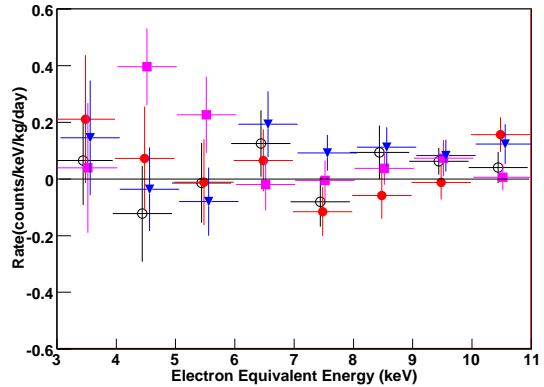}
\caption{\label{rate} (color online). Extracted NR event rates of the
S0501A~(open circles), S0501B~(filled circles), B0510A~(filled 
squares), and B05010B~(filled triangles) crystals and only 
statistical errors (1$\sigma$) are shown. The points are 
shifted with respect to each other on the $x$-axis to avoid overlapping.}
\vspace{-10pt}
\end{figure}

The estimation of the NR event rate is performed in each 1 keV bin 
from 3 to 11 keV for each crystal.
The $MT$ distributions of NR events and 
ER events are compared with the WIMP search data in Fig.~\ref{MT_comp} for the 5-6 keV energy 
range. The probability density functions~(PDF) for the ER and NR events are obtained by fitting 
these distributions. 
An unbinned maximum likelihood fit is performed with the $log(MT)$ 
distribution of the WIMP search data using the likelihood function,
\begin{eqnarray*}
\mathcal{L}_i= && {1\over n!}\times \exp\{-(N_{NR,i}+N_{ER,i})\} \\
 && \times \prod_{k=1}^n [ N_{NR,i}PDF_{NR,i}(x_k)+N_{ER,i}PDF_{ER,i}(x_k)], \\
\end{eqnarray*}
where the index $i$ denotes the $i$-th energy bin; $n=N_{NR,i}+N_{ER,i}$ 
is the total number of events; $N_{NR,i}$ and $N_{ER,i}$ are the numbers of 
NR and ER events, respectively; $PDF_{NR,i}$ and $PDF_{ER,i}$ 
are PDFs of NR and ER events, respectively; and 
$x_k = \log(MT)$ for each event.  
The NR event rates obtained for each bin and for each 
crystal after efficiency correction are shown in Fig.~\ref{rate}. The 
extracted NR event rates are consistent with a null 
observation of the WIMP signal.

In order to obtain the expected measured energy spectrum of a WIMP signal including 
instrumental effects, a Monte Carlo~(MC) simulation with GEANT4~\cite{geant} is used. 
A recoil energy spectrum is 
generated for each WIMP mass with the differential cross section, 
form factor, and quenching factor, as described in Ref.~\cite{smith}. 
The spin-dependent form 
factor for $^{133}$Cs calculated by Toivanen~\cite{Toivanen} 
is used, 
while for $^{127}$I, Ressell and Dean's calculation~\cite{ressel97} is 
used.
The photons generated with the fitted decay function
described above are propagated to the PMT and digitized in the same manner as in the experiment. 
Subsequently, the photoelectrons within given time windows are counted to check the trigger condition and to calculate energy.
In this manner, the trigger efficieny and energy resolution is  
accounted for in the expected energy spectrum.
The trigger efficiency is found to be higher than 99\%
above 3 keV. The simulation is verified with the energy 
spectrum obtained using 59.5 keV gamma rays from $^{241}$Am. The peak 
position and width of the distribution are very well reproduced for 
each crystal as described in Ref~\cite{kims06a}. 


\begin{table}
\caption{\label{tab:spinvalue}Spin expectation values for $^{133}$Cs and $^{127}$I}
\begin{ruledtabular}
\begin{tabular}{ccccc}
Isotope & J &$<S_p>$ &$<S_n>$ &Reference \\ 
\hline
$^{133}Cs$ & 7/2 & -0.370 & 0.003 & \cite{iachello} \\
$^{127}I$ & 5/2 &  0.309 & 0.075 & \cite{ressel97} \\
\end{tabular}
\end{ruledtabular}
\vspace{-10pt}
\end{table}

The total WIMP rate, $R$, for each WIMP mass is obtained by fitting the measured 
energy spectrum to the simulated one. 
The 90\% confidence level~(CL) limit on $R$ is calculated by the Feldman-Cousins's
approach in the case of Gaussian with a boundary at the origin~\cite{feldman} and then  
converted to the WIMP-nucleus cross section, $\sigma_{W-A}$.
Subsequently, the limits on WIMP-nucleon cross section is obtained from 
Ref.~\cite{smith}\cite{tovey} as follows:
$$\sigma_{W-n}=\sigma_{W-A}{\mu^2_n\over\mu^2_A}{C_n\over C_A},$$
where $\mu_{n,A}$ are the reduced masses of the WIMP-nucleon and 
WIMP-target nucleus of mass number $A$. ${C_A/{C_n}}=A^2$ for SI 
interactions and ${C_A/{C_n}}={4/3} \{a_p<S_{p}>+a_n<S_{n}>\}^2(J+1)/J$ for SD 
interactions. Here $a_p$, $a_n$ are WIMP-proton and WIMP-neutron SD 
couplings respectively.
The spin 
expectation values used for this analysis are shown in Table~\ref{tab:spinvalue}. 
Following the ``model-independent'' 
framework~\cite{tovey}, we report the allowed region in two cases for 
SD interaction: 
one for $a_n=0$, and the other for $a_p=0$. We express the 
WIMP-nucleon cross section as follows:
\begin{eqnarray*}
\sigma_{W-n}^{SI}&=&\sigma_{W-A}{\mu^2_n\over\mu^2_A}{1\over A^2}, \\
\sigma_{W-n,p}^{SD}&=&\sigma_{W-A}{\mu^2_{n,p}\over\mu^2_A}{3\over4}{J\over{(J+1)}}{1\over{<S_{n,p}>^2}},
\end{eqnarray*}
where we indicate pure proton~($p$, $a_n =0$) and pure neutron~($n$, 
$a_p = 0$) coupling for SD interaction. We also present the
allowed region in the $a_p - a_n$ plane with the following relation~\cite{tovey}:
$$
\left( \frac{a_p}{\sqrt{\sigma_{W-p}}} \pm 
\frac{a_n}{\sqrt{\sigma_{W-n}}} \right)^2 \leq \frac{\pi}{24G^2_F \mu_p^2},$$
where $G_F$ is the Fermi coupling constant.

The uncertainty in the $MT$ distribution results in the uncertainty of the NR event rate. 
The limited statistics of the calibration data and different crystals used for the neutron calibration 
and WIMP search data are 
the major sources of this uncertainty. The former is investigated by varying the fitted parameters in PDF function within errors.
The lattter is estimated by changing the mean of $MT$ by the difference between the crystals. 
The systematic uncertainties from these two souces are combined in quadrature resulting in 20-30\% of 
statistical uncertainties depending on the energy bins.
In addition, there are uncertainties 
in the MC estimation of the expected event rates due to the uncertainties 
in the quenching factors and the difference of energy resolution between 
the MC simulation and the data. The systematic error from the MC simulation is estimated 
to be 13.3\% of the limits. These systematic errors are combined with the 
statistical error in quadrature in the presented results.

\begin{figure}
\includegraphics[width=0.44\textwidth]{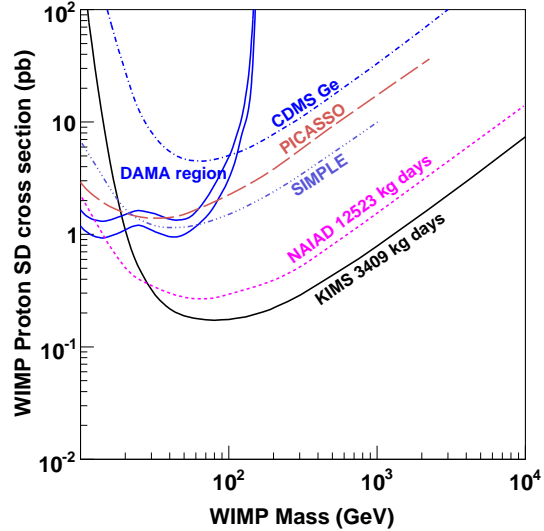}
\caption{\label{sdlimit_p}(color online). Exclusion plot for the SD interaction in the case of pure proton coupling~($a_n=0$) at the 90\% confidence level}
\vspace{-10pt}
\end{figure}
 
\begin{figure}[!]
\includegraphics[width=0.44\textwidth]{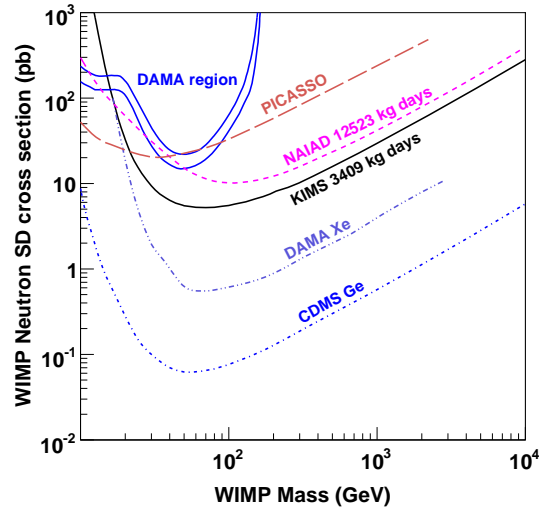}
\caption{\label{sdlimit_n} (color online). Exclusion plot for for the SD interaction in the case of pure neutron coupling~($a_p=0$) at the 90\% confidence level}
\vspace{-10pt}
\end{figure}

\begin{figure}
\includegraphics[width=0.44\textwidth]{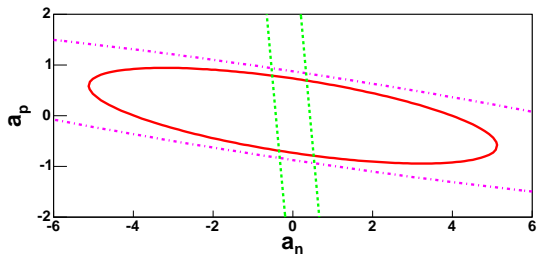}
\caption{\label{ap_an}(color online). Allowed region (90\% confidence level) in $a_p - a_n$ plane 
by KIMS data~(inside the solid line contour) for 50 GeV WIMP mass. Results of CDMS~\cite{akerib06}(dotted line) 
and NAIAD~\cite{naiad05}(dot-dashed line) are also shown.} 
\vspace{-10pt}
\end{figure}

\begin{figure}
\includegraphics[width=0.44\textwidth]{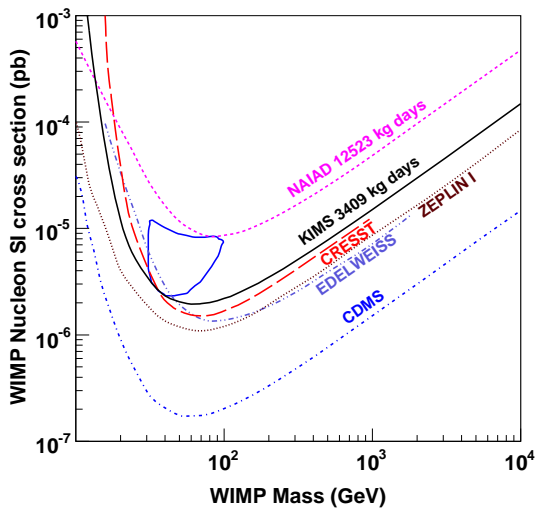}
\caption{\label{silimit}(color online). Exclusion plot for the SI interactions at the 90\% confidence level. 
} 
\vspace{-10pt}
\end{figure}

The limits on the SD interactions are shown in Fig.~\ref{sdlimit_p} and 
\ref{sdlimit_n} in the cases
of pure proton coupling and pure neutron coupling, respectively.
We also show the results obtained from CDMS~\cite{akerib06}, 
NAIAD~\cite{naiad05}, SIMPLE~\cite{simple05}, and 
PICASSO~\cite{picasso05}. The DAMA signal region is taken from 
Ref~\cite{savage04}. 
Our limit 
provides the lowest bound on 
the SD 
interactions in the case of pure proton coupling for a WIMP mass 
greater than 30~GeV/c$^2$. 
The allowed region in the $a_p - a_n$ plane for 
the WIMP mass of 50~GeV/c$^2$ is also shown in Fig.~\ref{ap_an} 
together with the 
limits from CDMS and NAIAD. 
The limit for the SI 
interactions is shown in Fig.~\ref{silimit} together with the results of 
CDMS~\cite{akerib06a}, EDELWEISS~\cite{edelweiss05}, 
CRESST~\cite{cresst05}, ZEPLIN I~\cite{zeplin05}, and the 3$\sigma$ signal region of 
DAMA (1-4)~\cite{bernabei00}. Although there are several experiments that 
reject the DAMA signal region, this is the first time that it is 
ruled out by a crystal detector containing $^{127}$I, which is the dominant nucleus for the SI interactions in the NaI(Tl) crystal.


In summary, we report new limits on the WIMP-nucleon cross section with CsI(Tl) 
crystal detectors using 3409~kg$\cdot$d exposure data. The DAMA signal regions for 
both SI and SD interactions are excluded for the WIMP masses higher than 20~GeV/c$^2$ by the single experiment. 
The most stringent limit on the SD interaction in the case of purely 
WIMP-proton coupling is obtained. 

The authors thank Dr. J. Toivanen and M. Kortelainen for the calculation of the SD form factor 
as well as for the useful discussions.
This work is supported by the Creative Research Initiative Program of 
the Korea Science and Engineering Foundation. We are grateful to the 
Korea Middland Power Co. Ltd. and the staff members of the YangYang Pumped 
Storage Power Plant for providing us the underground laboratory 
space. 


\end{document}